\documentclass[12pt]{article}
 \usepackage{graphicx,amsmath,amssymb}
 \usepackage[usenames]{color}
 \usepackage[colorlinks=true, urlcolor=navyblue, linkcolor=navyblue, citecolor=navyblue]{hyperref}
\definecolor{navyblue}{rgb}{0,0.08,0.45}

\newcommand{\mbf}[1]{\mathbf{#1}}
\newcommand{\half}{{\frac{1}{2}}}

\begin{document}

\begin{flushright}
{\small
SLAC--PUB--14526\\
\date{today}}
\end{flushright}

\vspace{30pt}

\centerline{\LARGE Excited Baryons in Holographic QCD}

\vspace{20pt}

\centerline{{
Guy F. de T\'eramond$^{a}$ and
%\footnote{Electronic address:gdt@asterix.crnet.cr}} 
Stanley J. Brodsky,$^{b,c}$ 
%\footnote{Electronic address:sjbth@slac.stanford.edu}
}}

\vspace{20pt}

{\centerline {$^{a}${Universidad de Costa Rica, San Jos\'e, Costa Rica}}

\vspace{4pt}

{\centerline {$^{b}${SLAC National Accelerator Laboratory, 
Stanford University, Stanford, CA 94309, USA}}

\vspace{4pt}

{\centerline {$^{c}${CP3-Origins, 
Southern Denmark University, Odense, Denmark}}

 \vspace{20pt}

\begin{abstract}
The light-front holographic QCD approach is used to describe baryon spectroscopy and the systematics of  nucleon transition form factors.
 \end{abstract}

\vspace{20pt}

\section{Introduction}

Baryon spectroscopy and the excitation dynamics of nucleon resonances encoded in the nucleon transition form factors  can provide fundamental insight into the strong-coupling dynamics of QCD.  The transition from the hard-scattering perturbative domain to the non-perturbative region is sensitive to the detailed dynamics of confined quarks and gluons. Computations of such phenomena from first principles in QCD are clearly very challenging.  The most successful theoretical approach thus far has been to quantize QCD on discrete lattices in Euclidean space-time;~\cite{Wilson:1974sk}
however, dynamical observables in Minkowski space-time, such as the time-like hadronic form factors are not amenable to Euclidean numerical lattice computations.

The AdS/CFT correspondence~\cite{Maldacena:1997re} between 
a gravity theory in anti--de Sitter (AdS) space-time and
conformal field theory in physical space-time  has   brought a new perspective
for the study  of strongly coupled quantum field theories.
A precise gravity dual to QCD is not known, but the mechanisms of
confinement can be incorporated into the gauge/gravity
correspondence by modifying the AdS$_5$ geometry in the  large
infrared  domain  $z \sim \frac{1}{\Lambda_{\rm QCD}}$, where 
$\Lambda_{\rm QCD} $ is the scale of the strong
interactions.~\cite{Polchinski:2001tt}  
The resulting dual theory incorporates an
ultraviolet conformal limit at the AdS boundary at $z \to 0$, as
well as the modifications of the background geometry in the large $z$
infrared region  which incorporate confinement. Despite its limitations, this approach to  the gauge/gravity duality, called AdS/QCD, has provided important physical  insights into the non-perturbative dynamics of hadrons -- in some cases, it is the only tool available.

The physics of AdS/QCD seems to be very abstract. However, ``Light-Front Holography'' leads to a rigorous connection between amplitudes in the fifth dimension of  AdS$_5$ space and physical $3+1$ space-time, thus providing a compelling physical interpretation of the AdS/CFT correspondence principle.
Light-front (LF)  holographic methods  were originally
introduced~\cite{Brodsky:2006uqa, Brodsky:2007hb} by matching the  electromagnetic  (EM) 
current matrix elements in AdS space~\cite{Polchinski:2002jw} with
the corresponding expression  using LF   theory in
physical space time.~\cite{Drell:1969km, West:1970av} It was also shown that one obtains  identical
holographic mapping using the matrix elements of the
energy-momentum tensor~\cite{Brodsky:2008pf} by perturbing the AdS
metric around its static solution.~\cite{Abidin:2008ku} One can also
study the AdS/CFT duality and its modifications starting from the Hamiltonian equation of motion
for a relativistic bound-state system
 $H^{QCD} _{LF}\vert  \psi(P) \rangle = P_\mu P^\mu \vert  \psi(P) \rangle = M^2 \vert  \psi(P) \rangle$
in physical space time,~\cite{deTeramond:2008ht} where 
the QCD light-front Hamiltonian $H^{QCD} _{LF} =P^+ P^- - \mbf{P}^2_\perp$, $P^\pm = P^0 + P^3$, is constructed from the QCD Lagrangian using the  standard methods of quantum field theory.~\cite{Brodsky:1997de}

To a first semiclassical approximation, where quantum loops and quark masses
are not included, LF holography leads to a LF Hamiltonian equation which
describes the bound-state dynamics of light hadrons in terms of
an invariant impact variable $\zeta$ which
measures the
separation of the partons within the hadron at equal light-front
time $\tau = x^0 + x^3$.~\cite{deTeramond:2008ht, Dirac:1949cp} In fact,  $\zeta^2 $  is related by a Fourier transform of the invariant mass of  the constituents, $M_n^2 = \left(\sum_{i=1}^n k_i^\mu\right)^2$, the key variable which controls the bound state.\footnote{For an $n$-parton hadronic system
the variable $\zeta$ is the $x$-weighted sum of the
transverse impact variables $\mbf{b}_{\perp i}$ of the $n-1$ spectator system:~\cite{Brodsky:2006uqa, Brodsky:2007hb}
$\zeta = \sqrt{\frac{x}{1-x}} ~ \Big\vert \sum_{j=1}^{n-1} x_j \mbf{b}_{\perp j} \Big\vert $,
where $x = x_n$ is the longitudinal
momentum fraction of the active quark. For a two-parton system $\zeta^2 = x(1-x) \mbf{b}_\perp^2$ and
$M^2_{q \bar q} =  \frac{\mbf{k}^2_\perp}{ x(1-x)}. $}

Remarkably, the AdS equations
correspond to the kinetic energy terms of  the partons inside a
hadron, whereas the interaction term  $U(z)$, build confinement and
correspond to the truncation of AdS space in an effective dual
gravity  approximation.~\cite{deTeramond:2008ht}
One also obtains a connection between the mass parameter $\mu R$ of the AdS theory with the orbital angular momentum of the constituents in the bound-state solutions of $H^{QCD} _{LF}$.
The identification of orbital angular momentum of the constituents is a key element in our description of the internal structure of hadrons using holographic principles, 
since hadrons with the same quark content, but different orbital angular momenta, have different masses.

\section{Nucleons in Light-Front Holography \label{NucleonSpec}}

For baryons, the light-front wave equation is a linear equation
determined by the LF transformation properties of spin 1/2 states. A linear confining potential
$U(\zeta) \sim \kappa^2 \zeta$ in the LF Dirac
equation leads to linear Regge trajectories.~\cite{Brodsky:2008pg}   For fermionic modes the LF matrix
Hamiltonian eigenvalue equation $D_{LF} \vert \psi \rangle = M \vert \psi \rangle$, $H_{LF} = D_{LF}^2$,
in a $2 \times 2$ spinor  component
representation, is equivalent to the system of coupled linear equations
\begin{eqnarray} \label{eq:LFDirac} \nonumber
- \frac{d}{d\zeta} \psi_- -\frac{\nu+\half}{\zeta}\psi_-
- \kappa^2 \zeta \psi_-&=&
M \psi_+, \\ \label{eq:cD2k}
  \frac{d}{d\zeta} \psi_+ -\frac{\nu+\half}{\zeta}\psi_+
- \kappa^2 \zeta \psi_+ &=&
M \psi_- ,
\end{eqnarray}
with eigenfunctions
\begin{eqnarray} \nonumber
\psi_+(\zeta) &\sim& z^{\frac{1}{2} + \nu} e^{-\kappa^2 \zeta^2/2}
  L_n^\nu(\kappa^2 \zeta^2) ,\\  \label{states}
\psi_-(\zeta) &\sim&  z^{\frac{3}{2} + \nu} e^{-\kappa^2 \zeta^2/2}
 L_n^{\nu+1}(\kappa^2 \zeta^2),
\end{eqnarray}
and  eigenvalues
%\begin{equation}
$M^2 = 4 \kappa^2 (n + \nu + 1) $.
%\end{equation}

The LF wave equation has also a geometric interpretation: it corresponds to the Dirac equation in AdS$_5$
space in presence of a linear potential $\kappa^2 z$
\begin{equation} \label{eq:DEz}
\left[i\big( z \eta^{M N} \Gamma_M \partial_N + 2 \, \Gamma_z \big) + \kappa^2 z
 + \mu R \right] \Psi = 0 ,
\end{equation}
as can be shown  by using the transformation $\Psi( z) \sim z^2 \psi(z)$, $z \to \zeta$.   The LF equation of motion  is thus mapped to a Dirac equation for spin-$\half$ modes  in  AdS space.

The baryon interpolating operator
$ \mathcal{O}_{3 + L} =  \psi D_{\{\ell_1} \dots
 D_{\ell_q } \psi D_{\ell_{q+1}} \dots
 D_{\ell_m\}} \psi$,  $L = \sum_{i=1}^m \ell_i$, is a twist 3,  dimension $9/2 + L$ with scaling behavior given by its
 twist-dimension $3 + L$. We thus require $\nu = L+1$ to match the short distance scaling behavior.   One can interpret $L$ as the maximal value of $|L^z|$ in a given LF Fock state.

In the case of massless quarks, the nucleon eigenstates have Fock components with different orbital angular momentum, $L = 0$ and $L = 1$, but  with equal probability.   {\it In effect, the nucleon's angular momentum is carried by quark orbital angular momentum. } Higher spin fermionic modes
 $\Psi _{\mu_1 \cdots \mu_{J-1/2}}$, $J > 1/2$, with all of its polarization indices along the $3+1$ coordinates follow by shifting dimensions for the fields as shown for the case of mesons in Ref.~\cite{deTeramond:2009xk}. 
Thus, as in the meson sector,  the increase  in the 
mass $M^2$ for baryonic states for increased radial and orbital quantum numbers is
$\Delta n = 4 \kappa^2$, $\Delta L = 4 \kappa^2$ and $\Delta S = 2 \kappa^2,$ 
relative to the lowest ground state,  the proton; i.e., the slope of the spectroscopic trajectories in $n$ and $L$ are identical .

\begin{figure}[!]
\begin{center}
\includegraphics[angle=0,width=14.0cm]{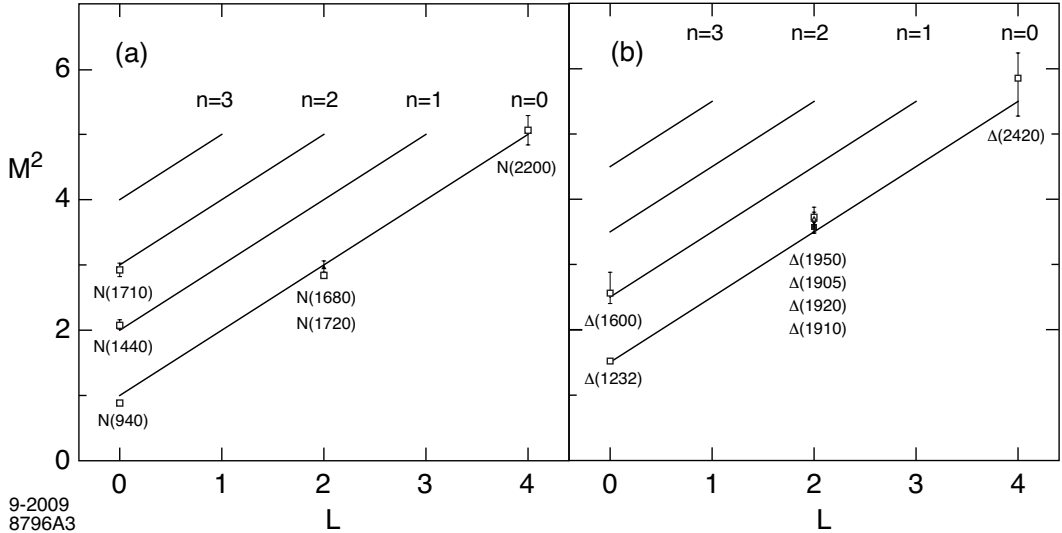}
\caption{Positive parity Regge trajectories for  the  $N$ and $\Delta$ baryon families for $\kappa= 0.5$ GeV}
\label{Baryons}
\end{center}
\end{figure}

We can fix the overall energy scale to be identical for mesons and baryons by imposing chiral symmetry for the pion~\cite{deTeramond:2010we} in the LF Hamiltonian equations: $m^2_\pi =0 $ for $m_q\to 0.$
The resulting predictions for the spectroscopy of positive-parity light baryons  are shown in Fig. \ref{Baryons}.
Only confirmed PDG~\cite{Amsler:2008xx} states are shown.
The Roper state $N(1440)$ and the $N(1710)$ are well accounted for in this model as the first  and second radial
states. Likewise the $\Delta(1660)$ corresponds to the first radial state of the $\Delta$ family. The model is  successful in explaining the parity degeneracy observed in the light baryon spectrum, such as the $L\! =\!2$, $N(1680)\!-\!N(1720)$ degenerate pair and the $L=2$, $\Delta(1905), \Delta(1910), \Delta(1920), \Delta(1950)$ states which are degenerate within error bars. The parity degeneracy of baryons is also a property of the 
hard-wall model, but radial states are not well described in this model.~\cite{deTeramond:2005su}  For other calculations of the baryonic spectrum in the framework of AdS/QCD, see Refs.~\cite{Hong:2006ta,  Forkel:2007cm, Nawa:2008xr,  Forkel:2008un, Ahn:2009px,  Zhang:2010bn, Kirchbach:2010dm, Gutsche:2011xx}.

An important feature of light-front holography is that it predicts the identical multiplicity of states for mesons
and baryons that is observed experimentally.~\cite{Klempt:2007cp} This remarkable property could have a simple explanation in the cluster decomposition of the
holographic variable, which labels a system of partons as an active quark plus a system on $n-1$ spectators. From this perspective, a baryon with $n=3$ looks in light-front holography as a quark--scalar-diquark system.

\section{Nucleon Form Factors}

In the higher dimensional gravity theory, hadronic amplitudes for the  EM transition $A\to B$ correspond to
the  non-local coupling of an external EM field $A^M(x,z)$  propagating in AdS with a  fermionic mode
$\Psi_P(x,z)$, given by the LHS of the equation below 
 \begin{multline} \label{FF}
 \int d^4x \, dz \,  \sqrt{g}  \,  \bar\Psi_{B, P'}(x,z)
 \,  e_M^A  \, \Gamma_A \, A_q^M(x,z) \Psi_{A, P}(x,z)  \\\sim
 (2 \pi)^4 \delta^4 \left( P'  \! - P - q\right) \epsilon_\mu  \langle \psi_B(P'), \sigma'  \vert J^\mu \vert \psi_A(P), \sigma \rangle,
 \end{multline} 
 where the coordinates of AdS$_5$ are the Minkowski coordinates $x^\mu$ and $z$ labeled $x^M = (x^\mu, z)$,
 with $M, N = 1, \cdots 5$, $g$ is the determinant of the metric tensor and $e^A_M$ is the vielbein with tangent indices
 $A, B = 1, \cdots, 5$.
The expression on the RHS  represents the QCD EM transition amplitude in physical space-time. It is the EM matrix element of the quark current  $J^\mu = e_q \bar q \gamma^\mu q$, and represents a local coupling to pointlike constituents.  Can the transition amplitudes be related for arbitrary values of the momentum transfer $q$? How can we recover hard pointlike scattering at large $q$ from the soft collision of extended objects?~\cite{Polchinski:2002jw}
Although the expressions for the transition amplitudes look very different, one can show that a precise mapping of the $J^+$ elements  can be carried out at fixed LF time, providing an exact correspondence between the holographic variable $z$ and the LF impact variable $\zeta$ in ordinary space-time.~\cite{Brodsky:2006uqa}

Hadronic form factors for the linear potential $\kappa^2 z$ discussed in Sec. \ref{NucleonSpec} -- for mesons the corresponding potential is a harmonic $\kappa^4 z^2$ ``soft wall'' potential~\cite{Karch:2006pv} -- have a simple analytical form.~\cite{Brodsky:2007hb} For a hadronic state with twist $\tau = N + L$  ($N$ is the number of components)
the form factor is expressed as a $\tau - 1$ product of poles along the vector meson Regge radial trajectory
 ($Q^2 = - q^2 > 0$)
\begin{equation} \label{F}   
 F(Q^2) =  \frac{1}{{\Big(1 + \frac{Q^2}{M^2_\rho} \Big) }
 \Big(1 + \frac{Q^2}{M^2_{\rho'}}  \Big)  \cdots 
       \Big(1  + \frac{Q^2}{M^2_{\rho^{\tau-2}}} \Big)} .
\end{equation}
For a pion, for example, the lowest Fock state -- the valence state -- is a twist-2 state, and thus the form factor is the well known monopole form.~\cite{Brodsky:2007hb}
The remarkable analytical form of Eq. (\ref{F}), expressed in terms of the $\rho$ vector meson mass and its radial excitations, incorporates the correct scaling behavior from the constituent's hard scattering with the photon and the mass gap from confinement.  It is also apparent from (\ref{F}) that the higher-twist components in the Fock expansion are relevant for the computation of hadronic form factors, particularly for the time-like region which is particularly sensitive to the detailed structure of the amplitudes.~\cite{deTeramond:2010ez}

Conserved currents are not renormalized and correspond to five dimensional massless fields propagating in AdS according to the relation 
$(\mu R)^2 = (\Delta - p) (\Delta + p -  4)$  for a $p$ form. In the usual AdS/QCD framework~\cite{Erlich:2005qh, DaRold:2005zs} this  corresponds to $\Delta = 3$ or 1, the canonical dimensions of
an EM current and the massless gauge field respectively.  Normally one uses a hadronic  interpolating operator  with minimum twist $\tau$ to identify a hadron in AdS/QCD and to predict the power-law fall-off behavior of its form factors and other hard 
scattering amplitudes;~\cite{Polchinski:2001tt} {e.g.},  for a two-parton bound state $\tau = 2$.   However, in the case of a current, one needs to  use  an effective field operator  with dimension $\Delta =3.$ The apparent inconsistency between twist and canonical dimension is removed by noticing that in the light-front one chooses to calculate the  matrix element of the twist-3 plus  component of the ``good" current  $J^+$,~\cite{Brodsky:2006uqa, Brodsky:2007hb} in order to avoid coupling to Fock states with different numbers of constituents.~\cite{Drell:1969km, West:1970av}

As mentioned above, light front holography provides a precise relation of the fifth-dimensional mass $\mu$ with the total  and orbital angular momentum of a hadron in the transverse LF plane $(\mu R)^2 = - (2 - J)^2 + L^2$, $L = \vert L^z\vert$,~\cite{deTeramond:2008ht} and thus  a conserved EM current  corresponds to poles along the $J=L=1$ radial  trajectory. For the twist-3 computation of the space-like form factor, which involves the current $J^+$, the poles do not correspond to the physical poles of the twist-2 transverse current $\mbf{J}_\perp$ present in the annihilation channel, namely the $J=1$, $L=0$ radial trajectory. Consequently, the location of the poles in the final result should be shifted to their physical positions. When this is done, the results agree extremely well with the proton Dirac elastic and transition form factor data shown in Fig. (\ref{pFFs}).
 
 \begin{figure}[!]
 \begin{center}
 \includegraphics[angle=0,width=7.6cm]{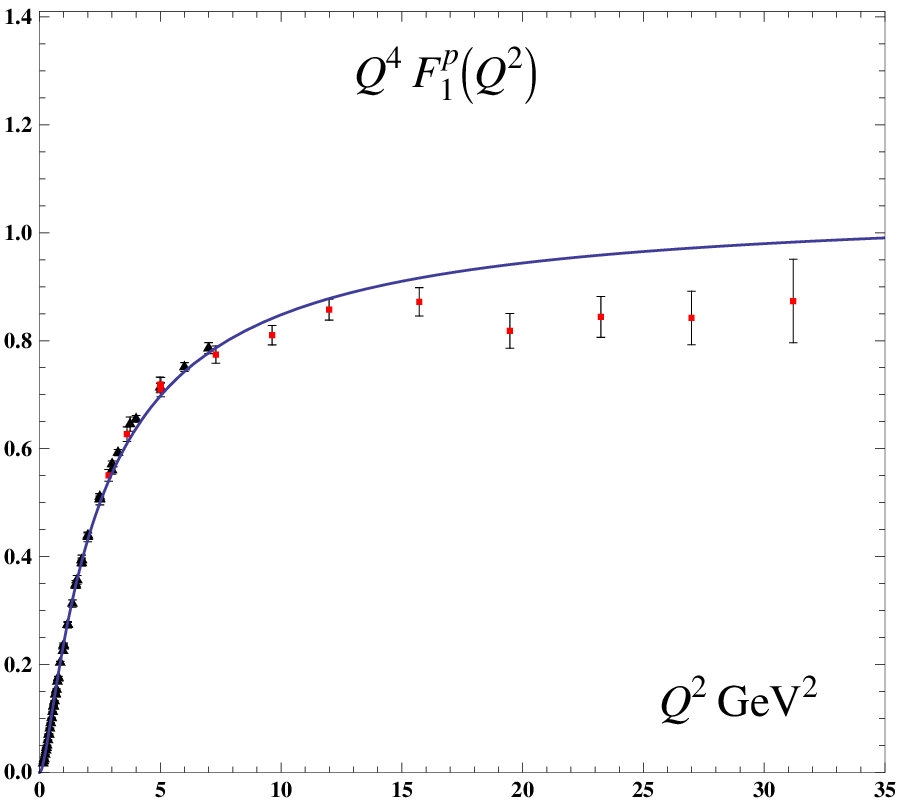}
\includegraphics[angle=0,width=7.6cm]{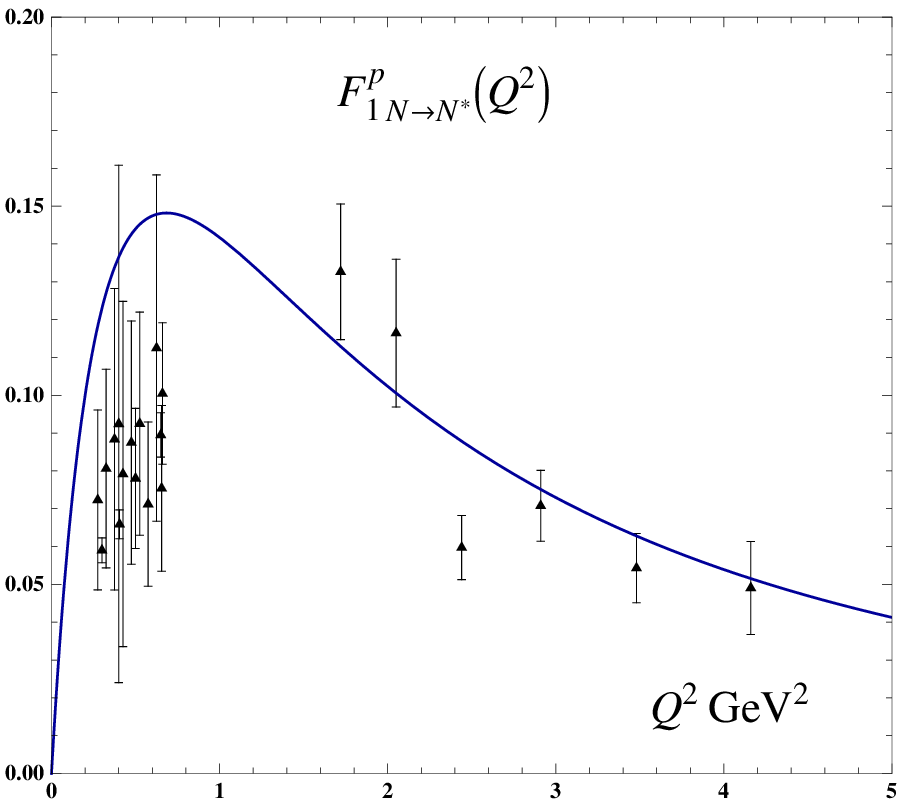}
\caption{Dirac proton form factors in light-front holographic QCD. Left: scaling of proton elastic form factor  $Q^4 F_1^p(Q^2)$. Right: proton transition form factor  ${F_1^p}_{N\to N^*}(Q^2)$ to the first radial excited state. Data compilation  from Diehl~\cite{Diehl:2005wq} (left) and JLAB~\cite{Aznauryan:2009mx}  (right).}
\label{pFFs}
\end{center}
\end{figure}

The proton has degenerate mass eigenstates with plus and minus components (\ref{states})
corresponding to $L^z=0$ and $L^z= + 1$ orbital components combined with spin components $S^z = + 1/2$ and $S^z = -1/2$ respectively. Likewise, we would expect that the wave equation describing the propagation of a vector meson in AdS with $J_z = + 1$ will account for three degenerate mass eigenstates with different LF orbital angular momentum components:  $L^z = 0$, $S^z = + 1$;  $L^z = +1$, $S^z = 0$ and $L^z = +2$, $S^z = -1$, which is obviously not the case in the usual formulation of AdS wave equations. To describe higher spin modes in AdS/QCD, properly incorporating the spin constituents,
the formalism  has to be extended to account for multiple component wave equations with degenerate mass eigenstates -- as for the case of the nucleon --  introducing coupled linear equations in AdS similar to the  Kemmer-Duffin-Petiau equations,  a subject worth pursuing. 

\subsection{Computing Nucleon Form Factors in Holographic QCD}

In order to compute the separate features of the proton an neutron form factors  one needs to incorporate the spin-flavor structure of the nucleons,  properties  which are absent in the usual models of the gauge/gravity correspondence.
This can be readily included in AdS/QCD by weighting the different Fock-state components  by the charges and spin-projections of the quark constituents; e.g., as given by the $SU(6)$  spin-flavor symmetry. 

To simplify the
discussion we shall consider the spin-non flip proton form factors for the transition $n,L \to n' L$. Using the $SU(6)$ spin-flavor symmetry we obtain the result~\cite{Brodsky:2008pg} 
\begin{equation} \label{F1}
{F_1^p}_{n, l \to n', L}(Q^2)  =    R^4 \int \frac{dz}{z^4} \, \Psi_+^{n' \!, \,L}(z) V(Q,z)  \Psi_+^{n, \, L}(z),
\end{equation}
where we have factored out the plane wave dependence of the AdS fields
\begin{equation} \label{Psip}
\Psi_+(z) = \frac{\kappa^{2+L}}{R^2}  \sqrt{\frac{2 n!}{(n+L+1)!}} \,z^{7/2+L} L_n^{L+1}\!\left(\kappa^2 z^2\right) 
e^{-\kappa^2 z^2/2}.
\end{equation}
The bulk-to-boundary propagator
is~\cite{Grigoryan:2007my}
\begin{equation} \label{V}
V(Q,z) = \kappa^2 z^2 \int_0^1 \! \frac{dx}{(1-x)^2} \, x^{\frac{Q^2}{4 \kappa^2}} 
e^{-\kappa^2 z^2 x/(1-x)} ,
\end{equation}
with $V(Q = 0, z) = V(Q, z = 0) =1$. The orthonormality of the Laguerre polynomials in (\ref{Psip}) implies that the nucleon form factor at $Q^2 = 0$ is one if $n = n'$ and zero otherwise. Using  (\ref{V}) in (\ref{F1}) we find
\begin{equation} \label{protonF1}
F_1^p(Q^2) =  \frac{1}{{\Big(1 + \frac{Q^2}{M^2_\rho} \Big) }
 \Big(1 + \frac{Q^2}{M^2_{\rho'}}  \Big) },
 \end{equation}
 for the elastic proton Dirac form factor and
 \begin{equation} \label{RoperF1}
 {F_1^p}_{N \to N^*}(Q^2) = \frac{\sqrt{2}}{3} \frac{\frac{Q^2}{M^2_\rho}}{\Big(1 + \frac{Q^2}{M^2_\rho} \Big) 
 \Big(1 + \frac{Q^2}{M^2_{\rho'}}  \Big)
       \Big(1  + \frac{Q^2}{M^2_{\rho^{''}}} \Big)},
\end{equation}
for the EM spin non-flip proton to Roper  transition form factor. The results (\ref{protonF1}) and (\ref{RoperF1}),
 compared with available data in Fig. \ref{pFFs}, correspond to the valence approximation.   The transition form factor
 (\ref{RoperF1}) is expressed in terms of  the mass of the $\rho$ vector meson and its first two radial excited states, with no additional parameters.
 
To study the spin-flip nucleon form factors using holographic methods, Abidin and Carlson~\cite{Abidin:2009hr} 
 propose to 
introduce the `anomalous' gauge invariant term 
\begin{equation}
\int d^4x~dz~\sqrt{g}  ~ \bar\Psi
 \,  e_M^A\,  e_N^B \left[\Gamma_A, \Gamma_B\right] F^{M N}\Psi
 \end{equation}
 in the five-dimensional action, since the structure of (\ref{FF}) can only account for $F_1$.  Although this is a practical  
 avenue, the overall strength of the new term has to be fixed by the static quantities and thus some predictivity is lost. We hope that further progress using light-front holographic methods will overcome this shortcoming as well as the other difficulties described in this article.
 
 Holographic QCD methods have also been used to obtain general parton distributions (GPDs) in
 Refs.  \cite{Vega:2010ns} and \cite{Nishio:2011xa},  and a study of the EM nucleon to $\Delta$ transition form factors  has been carried out in the framework of the Sakai and Sugimoto model in Ref. \cite{Grigoryan:2009pp}.

\section{Conclusions}

Light-front holography provides a simple and successful framework for describing the  systematics of the excitation spectrum of baryons: the mass eigenspectrum,  observed multiplicities and degeneracies. It also provides new analytical tools  for computing hadronic transition amplitudes, incorporating the scaling behavior and the transition to the confinement region.  The framework has a simple analytical structure and can be applied to study dynamical properties in Minkowski space-time which are not amenable to Euclidean  lattice gauge theory computations. The framework  for higher-spin hadrons can be improved by allowing for the multi-component structure of wave equations. A fully comprehensive framework should also include the spin-flavor symmetry in  the gauge/gravity correspondence.  In spite of its present limitations,  the AdS/QCD approach, together with light-front holography, provides  important physical  insights into the non-perturbative regime of QCD and its transition to the perturbative domain where quark and gluons are the relevant degrees of freedom. The new set of tools provided by the gauge/gravity correspondence and light front holography will thus be useful for the theoretical interpretation of the results to be obtained at the new mass scale and kinematic regions accessible to the JLab 12 GeV Upgrade Project.

\section*{Acknowledgments}

Invited talk presented by GdT at NSTAR 2011, the 8th International Workshop on the Physics of Excited Nucleons,
Jefferson Laboratory, May 17 - 20,  2011. GdT  is grateful to the organizers for their  hospitality at JLAB. We thank our collaborators F. G. Cao, H. G. Dosch and J. Erlich for many helpful conversations. This research was supported by the Department of Energy  contract DE--AC02--76SF00515.

\end{document}